%% file: DSC_IEEE.tex
\RequirePackage{fix-cm}
\RequirePackage{snapshot}
\documentclass[journal]{IEEEtran}
\ifCLASSINFOpdf
\else
\fi
\usepackage[inline]{enumitem} 
\setlist*[enumerate,1]{%
  label=(\roman*),
}
\usepackage{diagbox}
\usepackage{cite}
\usepackage[table]{xcolor}
\usepackage{pgfplots}
\usepackage{pgfplotstable}
\usepackage{xparse}

\usepackage{wrapfig}

\usetikzlibrary{matrix}
\usepgfplotslibrary{groupplots}
\pgfplotsset{compat=newest}
\usetikzlibrary{positioning}

\usepackage{smartdiagram}
\pgfplotsset{compat=1.5}
\usepackage{amssymb, amsmath}

\usepackage{caption}
\usepackage{subcaption}
\usepackage[hyphens]{url}
\usepackage{epstopdf}
\usepackage{algorithm}
\usepackage{algpseudocode}
\usepackage{expl3}
\usepackage{url}
\usepackage{lipsum}

\input{macros.tex}

\begin{document}

\title{Robust Machine Learning for Encrypted Traffic Classification}
 
\author
{
		\IEEEauthorblockN
		{
			Amit Dvir\IEEEauthorrefmark{1}\IEEEauthorrefmark{3},
			Yehonatan Zion\IEEEauthorrefmark{1}\IEEEauthorrefmark{3},
			Jonathan Muehlstein\IEEEauthorrefmark{1}\IEEEauthorrefmark{3}, 
			Ofir Pele\IEEEauthorrefmark{2}\IEEEauthorrefmark{3}
			Chen Hajaj\IEEEauthorrefmark{3}\IEEEauthorrefmark{4}
			Ran Dubin\IEEEauthorrefmark{3}, \\
		}
		\IEEEauthorblockA
		{
			\IEEEauthorrefmark{1}
			Department of Computer Science, Ariel University, Israel \\
		}
		\IEEEauthorblockA
		{
			\IEEEauthorrefmark{2}
			Department of Electrical and Electronics Engineering, Ariel University, Israel\\
		}
		\IEEEauthorblockA
		{
			\IEEEauthorrefmark{3}
			Ariel Cyber Innovation Center, Ariel University, Israel\\
		}
		\IEEEauthorblockA
		{
			\IEEEauthorrefmark{4}
			Department of Industrial Engineering \& Management, Ariel University, Israel\\
		}

} 

\maketitle  

\begin{abstract}
Desktops and laptops can be maliciously exploited to violate privacy. In this paper, we consider the daily battle between the passive attacker who is targeting a specific user against a user that may be adversarial opponent. In this scenario, while the attacker tries to choose the best vector attack by surreptitiously monitoring the victim's encrypted network traffic in order to identify user's parameters such as the Operating System (OS), browser and apps. The user may use tools such as a Virtual Private Network (VPN) or even change protocols parameters to protect his/her privacy. We provide a large dataset of more than 20,000 examples for this task. We run a comprehensive set of experiments, that achieves high (above $85\%$) classification accuracy, robustness and resilience to changes of features as a function of different network conditions at test time. We also show the effect of a small training set on the accuracy. 
\end{abstract}


\section{Introduction}
Today, the amount of sensitive data that users handle with devices is growing rapidly. Typical users access their bank web-site, social network web-sites and get information about diseases, sexual or religious preferences that are assumed to be confidential and non-hackable.

Personal devices can be attacked in two broad ways which are known as passive and active. Active adversaries try to physically or remotely control the
user's device. Passive adversaries may violate the privacy of the user by sniffing the network traffic of the devices from the network side. A passive adversary can use this information to map user actions and obtain information on personal interests and habits. If the network traffic is not encrypted, a passive attacker can simply analyze the unencrypted HTTP requests and responses \cite{User_web,OSN_12,FFA_09,AtenieseHMVV15}. However, most Internet traffic today is encrypted as a result of increased user awareness of privacy threats and encouragement by Google to get all website owners to switch from HTTP to HTTPS by indicating whether sites use secure, encrypted connections in Google ranking algorithms \cite{GoogleSSL,TLS}. One immediate outcome has been that traditional Deep Packet Inspection (DPI) methods for information retrieval are no longer viable.

Nevertheless, numerous studies have shown that encryption is not sufficient both in desktop and mobile   \cite{AppScanner,M_actions,CGPS06,SSHAKT07,white2011phonotactic,wright2007language,dainotti2012issues,alshammari2010unveiling,zander2005self,paredes2012practical,zhang2010identification,bonfiglio2009detailed,chen2006quantifying,bar2010realtime,valenti2013reviewing,cao2014survey,BCB15_1,NA08_8,BLPR10_2,NR14_12,LychevJBN15_SecureQuic,SSL_Peek,Shi20171,SSLFeat}.
Bujlow et al. \cite{BCB15_1} presented a survey on popular DPI tools for
traffic classification.  Nguyen and Armitage \cite{NA08_8} surveyed
machine learning techniques for Internet traffic classification.
Moore et al. Niemczyk et al. \cite{NR14_12} suggested dividing the session into time buckets (10 seconds). They conclude that the new feature (buckets) was almost perfect for Skype but not able to differentiate between browsers and between joint application and browser usage.

In this work, we consider the daily battle between the passive attacker who is targeting a specific user which in some cases the user may be adversarial opponents (e.g., \cite{HJNRT11}). In this scenario, while the attacker tries to choose the best vector attack against the user by surreptitiously monitoring the victim's network traffic to identify users parameters such as the Operating System (OS), browser and apps, the user may use tools such as a Virtual Private Network (VPN) or even change protocols parameters to protect his/her privacy. The attacker can either sniff the wireless network of the user or can sniff the user traffic over the ISP network (e.g., government). Based on the vulnerabilities of the tuple (OS, Browser, Applications), the attacker will choose the best malwares or Advanced Persistent Threat (APT) method and due to the encryption the attacker may consider using machine learning to classify those user parameters (OS, browser and applications). 

Although studies such as the above have shown that encryption is not sufficient and presented systems to identify and classify user parameters such as the operating system (e.g. \cite{P0f,OS_new,Husk2016}), the user browser (e.g. \cite{Husk2016}) and the user applications (e.g. \cite{skype_4, Alshammari_7}). None of them have dealt with the case where the user may be aware of the classification system and decides to become an adversarial opponent. In this case, the user will try to manipulate the data to protect privacy by changing the network traffic or the protocols parameters. Due to the fact that the passive attacker may use machine learning algorithms, the user may change his traffic during testing time and not in training which increases the difficultly of the passive attacker to classify the user parameters. 

To respond to this need we first present a machine learning algorithms, for the passive attacker, that uses new features to classify the user tuple (OS, Browser, Application). Second, we show the performance of our machine learning (robustness) to attacks by adversarial opponent that implement tools such as VPN which extends a private network. Third, we show that our system is also robust against changes in protocol parameters (e.g. cipher suite).
The last two achievements are the main contribution of the paper. 

Note that, an example for the motivation of the attacker of knowing the operation system, browsers can be seen in vulnerability CVE-2019-5786 \cite{CVE} where the application combination can be seen in \cite{flash}. Knowing the tuple give us the ability to selective our exploits which will be specify for the specific user.

Finally, we show that our algorithm achieves high accuracy even when using a small number of training samples or taking only a short session time as a test sample. 

In summary:

\begin{itemize} 
\item We present a machine learning system to identify the user's operating system, browser and application from HTTPS traffic which achieves 96.06\% accuracy. Our machine learning use new features that exploit browsers' bursty behavior and SSL behavior.
	
\item Our system only incurs marginal damage in the case of adversarial opponent. Where the adversarial opponent may change the protocol parameters, use network tools or shortage the information (small amount of session packets). We show that our solution is robust to changes in cipher suites (protocol changes), 94\%, and VPN (network tool), 83\%, where both changes are \emph{only} on the testing data (i.e. the user changed the parameters whereas the machine learning training and classification did not change). 

\item We also show that using small training datasets only incurs marginal damage. For example, using only $500$ examples for training (vs. the full training dataset of 0.7$\times$20,633) we achieved a reasonable accuracy of 85\%.

\item We investigate the resilience of the system to short session times; for example, using only the first $10$ seconds of a session for training and testing. We show that although we use shorter sessions which yields less meta data, we achieve 94.2\% accuracy. Note that, short session times show that our system can be a real time system (do not need to wait to the end of the session).

\item We provide a comprehensive dataset that contains more than 20,000 labeled sessions. The operating systems tested were: Windows, Linux-Ubuntu and OSX. The browsers tested were: Chrome, Internet Explorer, Firefox and Safari. The applications tested some over browsers and some standalone were: YouTube, Facebook, Twitter, Teamviewer and Dropbox \footnote{For the sack of simplicity, we define Youtube, Facebook, Twitter over the browser as an application. The motivation based on the fact that the common access way over a desktop is using the browser}. The dataset is available for download at \cite{RDS} and the codes for the crawler, feature extractions and machine learning algorithms can be found in \cite{OurCrwalerCode,OurFeaturesCode,OurMLCode}. 
\end{itemize}

In this paper, we extend our previous work \cite{BOA_conf} on machine learning algorithms that classify tuples. In this work, we investigated the robustness and resilience of our system against adversarial opponents users. These users attempting to protect their privacy and are aware of the fact that their privacy may been infringed. We also introduce two machine learning algorithms and extended datasets and discuss possible countermeasures of our system.

The remainder of this paper is organized as follows In Section \ref{related_work} we review the state of the art on this topic. Section \ref{BOA} presents our solution to identifying the user's operating system, browser and application. In Section \ref{Results} we evaluate our method including its robustness to changes in network conditions at test time.  In Section \ref{Possible Countermeasures and Limitations} we discuss limitations and possible countermeasures. Finally, we discuss future lines of work in Section \ref{Conclusions and Future Work}.

\section{Related Works}
\label{related_work}  
Feature extraction methods for traffic classification, application classification and application analysis include session
duration \cite{Alshammari_7}, number of packets in a session
\cite{skype_4,Oliva_5}, minimum, maximum and average values of
inter-arrival packets time \cite{skype_4, Alshammari_7}, payload size
information \cite{skype_4}, bit rate \cite{Chen_16, Bonfiglio_17},
round-trip time \cite{Chen_16}, packet direction \cite{Hjelmvik_18}, SSL parameters \cite{Shi20171}
and server sent bit-rate \cite{Bar_19}. 

Liberatore and Levine \cite{liberatore2006inferring} showed the effectiveness of two traffic analysis techniques for the identification of encrypted HTTP streams. One is based on a \naive{} Bayes classifier and the other on Jaccard's coefficient similarity measure. They also proposed several methods for actively countering these techniques. They found these methods to be effective, albeit at the cost of a significant increase in the size of the traffic stream.  Panchenko \etal \cite{panchenko2011website} showed that a Support Vector Machine (SVM) classifier is able to correctly identify web pages, even when the user used both encryption and anonymization networks such as Tor \cite{tor}. Cai \etal \cite{cai2012touching} presented a web page
fingerprinting attack and showed that it is able to overcome defenses such as the application-level defense HTTPOS \cite{luo2011httpos} and
randomized pipelining over Tor.

Mobile devices, which have different operating systems and different applications implementation leading in many cases to different
network behavior, are also susceptible to attackers seeking information on user privacy, applications \cite{ILDM,TaintDroid,smartphone_traffic,Mobile_sync,mobile_ProfileDroid,sinha2016flowmine,fereidooni2016efficient} and actions \cite{mobile_Saltaformaggio,M_actions,M_actions_conf}. Saltaformaggio
et al. \cite{mobile_Saltaformaggio} presented NetScope, a passive framework for identifying user activities within the wireless network
traffic based on inspecting IP headers. Conti et al. \cite{M_actions,M_actions_conf,AppScanner,conti_new_2017} devised a highly accuracy classification frameworks for various user mobile actions and applications using network features such as size, direction (incoming/outgoing), and timing.

Gathering information on the Operating System (OS) of the user can be useful too. Passive sniffing of the OS fingerprinting
techniques was proposed in \cite{NetRESec,P0f,OS_class,OS_new}. p0f \cite{P0f} is a well-known and widely used tool that uses header fields such as the TCP SYN and SYN-ACK exchanges to fingerprint the OS. However, any change in the 3-way handshake of the TCP will affect the tool. Anderson and McGrew \cite{OS_new}  presented an effective approach to OS passive fingerprinting that uses a combination of encrypted data (TLS+TCP/IP) with non-encrypted data (HTTP) over multi-sessions. Although certain papers \cite{P0f,OS_new} classified the operating system, they based their classification on either protocol parameters (TCP handshake) \cite{P0f} or on non-encrypted data \cite{OS_new}. Moreover, none of them considered the case of adversarial opponents that try to protect against the classification.

Passive fingerprinting of browser clients from encrypted traffic is also necessary. Husak et al. \cite{Husk2016} proposed real-time exact pattern matching for the identification of the user's OS or browser based on SSL/TLS fingerprinting. In this work\cite{Husk2016}, the system has to identify the SSL parameters and is not robust to changes in the SSL parameters such as cipher suite parameters by adversarial opponents (the user). 

Machine learning systems offer unparalled flexibility in dealing with evolving input in a variety of applications. However, machine learning algorithms themselves can be a target of attack by a malicious adversary. Barreno et al. \cite{isSecure} described different types of attacks on machine learning and a variety of defenses against these attacks. Demontis et al. \cite{yesML} presented a simple and scalable secure-learning paradigm that mitigates the impact of invasion attacks, while only slightly affecting the detection rate. In our paper attacks on machine learning can be by using network tools such as VPN or changing the protocol parameters.

\section{Robust Identification of Users' Operating System, Browser and Application}
\label{BOA}
The goal of this paper is to identify the user operating system, browser and
application even in the case where the user is aware of the system and may try to decrease the accuracy of the system by changing protocol parameters and/or by using network tools. To achieve this goal, we used supervised machine learning
techniques. 

Supervised machine learning techniques learn a function which given a sample returns a label. Learning is carried out using a dataset of labeled samples. In our case, we chose to use sessions as samples, where a session is the tuple \textless Protocol, IP source, IP destination, Port source, Port destination\textgreater{} and the label is the tuple \textless OS, Browser, Application\textgreater. Thus, our task is inherently a form of multiclass learning with more than $30$ classes (see Figure \ref{mp:tuplestat} for the labels and their statistics in the dataset).

The rest of this section is organized as follows In section
\ref{sec:Dataset} we outline how we collected the dataset and the
dataset characteristics. In section \ref{sec:Feature Extraction} we
describe and discuss our feature extraction scheme.  Finally, in
section \ref{sec:Learning} we provide details on the machine learning
methodology we used.

\subsection{Dataset}
\label{sec:Dataset}
In order to simulate different usage scenarios, we built an automatic system that emulates user browsing (websites and associated actions). We used the Selenium web automation tool \cite{Selenium} to develop crawlers to gather the dataset. We collected all the traffic that passed through port $443$ (TLS/SSL). Finally, we split the traffic into sessions using SplitCap \cite{SplitCap}. 

We used the crawlers on a standard Internet connection over various operating systems and various browsers and combinations thereof. The passive users follow the active user and all hosts capture traffic. The dataset was collected automatically over the period of more than two months in our research lab over diverse connections (wired and WiFi) and networks conditions (over workdays and weekends, 24/7). The lab architecture contain $8$ computers, with various operation systems and browsers where each entity ran a crawler; the crawler code can be found in \cite{OurCrwalerCode}.

The crawler covered multiple sequences of UI actions across different systems and configurations. This presented various challenges. First, we were faced with an unclear workflow when we had to add features to the code base. To handle this issue we defined a configuration file with all the required simulation information. The configuration file was stored in a JSON file format. This method allowed us to abstract the core functionality code from the configuration of each computer and made it easier to handle bugs and add features.

Second, the crawler had to work on several different applications and functions both as an active user and as well as a passive user. To do so, we made the crawler generic such that for each application all we needed to do was add the XPATHs (XPATH is a query language for node selection from within an XML document.) of the fields of each action and run it. This idea was good in theory but involved technical complexity that wasted too much time. Thus, we developed a crawler for each application. 

Third, the crawler had to handle application updates both in the context of security as well as the UI format. It makes sense that services like YouTube, Facebook and Twitter will change the UI of their services. It is also reasonable that from time to time they will change their privacy and security policy. Changes such as these gave us considerable trouble since each change in the UI affected the HTML design and hence hindered our crawler's effectiveness since the crawler relies on the HTML structure. 

Finally, we had to deal with the way web services handle bots like our own. To prevent abusive behavior, social media platforms enforce timing constraints such as $X$ posts per day and a single new post per Y seconds. To overcome this issue we experimented with multiple timing configurations.

For Facebook, the same account was used for both sending and receiving posts. For Twitter, we had one sending account and several receiving accounts (followers) which ranged over various operating systems and various browsers and combinations thereof. Teamviewer traffic was generated by us actively without a crawler. 
In addition to our active traffic, we also observed background traffic that the operating systems, browsers and applications created (Google-Services, Microsoft-Services). One example of a service is Google Analytics or Microsoft Live. Dropbox traffic was composed both of active (no crawler) and background traffic. Any traffic that we could not identify was labeled unidentified. The browser label part of the tuple of stand-alone applications which do not work under a browser (e.g., Dropbox, Teamviewer) were labeled as Non-Browser. 

Thus overall our dataset contained more than $20,000$ sessions. The average duration of a session was $518$ seconds where on average each session had $520$ forward packets (the average forward traffic size was $261$ Kbytes) and $637$ backward packets (average backward traffic size is $615$ Kbytes). The tuple label statistics can be seen in Figure \ref{mp:tuplestat}. The operating system, browser and application statistics can be seen in Figures \ref{mp:osstat},\ref{mp:browswerstat},\ref{mp:appstat} respectively. The dataset can be found in \cite{RDS}. 

\input{stat_fig.tex}

\subsection{Feature Extraction}
\label{sec:Feature Extraction}
Using raw data in a machine learning method is problematic because the data are non-structured and contain redundant information. Thus, there is a need to build a structured representation of the raw data that is informative as to the specific problem domain. Building this representation is called feature
extraction \cite[Chapter~5.3]{friedman2001elements}.

We extracted features from a session of encrypted traffic which generally relies on SSL/TLS for secure communication. These protocols are built on top of the TCP/IP suite. The TCP layer receives encrypted data from the above layer and divides the data into chunks if the packets exceed the Maximum Segment Size (MSS). Then, for each chunk it adds a TCP header creating a TCP segment. Each TCP segment is encapsulated into an Internet Protocol (IP) datagram. As TCP packets do not include a session identifier, we identified a session using the tuple \textless Protocol, IP source, IP destination, Port source, Port destination\textgreater.

\begin{table}[t!]
  \begin{minipage}{0.8\linewidth}
  \centering
  \begin{tabular}{|l|}
	\hline
    \# Forward packets \\\hline
	\# Forward total bytes\\\hline
	Min forward inter arrival time difference\\\hline
	Max forward inter arrival time difference\\\hline
	Mean forward inter arrival time difference\\\hline
	STD forward inter arrival time difference\\\hline
	Mean forward packets\\\hline
	STD forward packets\\\hline
	\# Backward packets\\\hline
	\# Backward total bytes\\\hline
	Min backward inter arrival time difference\\\hline
	Max backward inter arrival time difference\\\hline		
	Mean backward inter arrival time difference\\\hline
	STD backward inter arrival time difference\\\hline
	Mean backward packets\\\hline
	STD backward packets\\\hline
	Mean forward TTL value \\\hline
	Minimum forward packet\\\hline
    Minimum backward packet\\\hline
    Maximum forward packet\\\hline
    Maximum backward packet\\\hline
    \# Total packets\\\hline
    Minimum packet size\\\hline
    Maximum packet size\\\hline
    Mean packet size\\\hline
    Packet size variance\\\hline
  \end{tabular}
  \subcaption{\baseset}
  \label{tab:basefeatures}
  \vspace{0.5cm}
  \begin{tabular}{|l|}
	\hline
    TCP initial window size\\\hline
    TCP window scaling factor\\\hline		
    \# SSL compression methods\\\hline
    \# SSL extension count\\\hline
	\# SSL cipher methods\\\hline
	SSL session ID len\\\hline
	Forward peak MAX throughput\\\hline
	Mean throughput of backward peaks\\\hline		
	Max throughput of backward peaks \\\hline
	Backward min peak throughput\\\hline
	Backward STD peak throughput\\\hline
	Forward number of bursts\\\hline
	Backward number of bursts\\\hline
	Forward min peak throughput\\\hline
	Mean throughput of forward peaks\\\hline
	Forward STD peak throughput\\\hline
	Mean backward peak inter arrival time diff\\\hline
    Minimum backward peak inter arrival time diff\\\hline
    Maximum backward peak inter arrival time diff\\\hline
    STD backward peak inter arrival time diff\\\hline
    Mean forward peak inter arrival time diff\\\hline
    Minimum forward peak inter arrival time diff\\\hline
    Maximum forward peak inter arrival time diff\\\hline
    STD forward peak inter arrival time diff\\\hline
    \# Keep alive packets\\\hline
	TCP Maxiumu Segment Size\\\hline
    Forward SSL Version\\\hline
  \end{tabular}
  \subcaption{\newset}
  \label{tab:newfeatures}
  \end{minipage}
  \vspace{0.1cm}
  \caption{The \baseset{} used in previous traffic classification methods and the \newset{} proposed in this paper.}
  \label{tab:allfeatures}
\end{table}

A session contains two flows: forward and backward. A flow is defined as a time ordered sequence of TCP packets during a single TCP session. The forward flow is defined as a time series of bytes transported by incoming packets alone, whereas the backward flow is defined as a time series of bytes transported solely by outgoing packets. We used the forward and backward flows and their combination as a representation of a connection. Additionally, we also used time series features such as the inter-arrival time differentials between different packets on the same flow. Based on our previous work \cite{dubin2016know}, for classifying video titles, we also used the bursty behavior of the browsers (peaks) which is defined as a section of traffic preceded and followed by silence. An example of the bursty behavior of browsers is depicted in Figure \ref{fig:chrome_720_auto_costarica_wireshark}. Note that the bursty behavior of browser traffic was also observed for YouTube traffic in \cite{PRNLJ_12,AALTBD_11}. The feature extraction takes the session network traffic as input and extracts features from it. 

In the results section we show that this combination (named as Combined set) considerably outperforms previous works (base-line set, termed the common set). We consider nine sets of features as can be seen in Table \ref{tab:FS}. The feature extraction code can be found in \cite{OurFeaturesCode}.

\begin{table}
\begin{tabular}{|p{2.5cm}|p{5.5cm}|}
    \hline
		  Common Feature Set& a set of features used in previous traffic classification methods \cite{Williams_9,NR14_12,McGregor_11,Moore_6,Moore_10,Alshammari_7,OS_class} (presented in Table \ref{tab:basefeatures}). This feature set includes the common features in previous papers which we used as our baseline, and as a comparison to our features set. 
\\\hline 
      Peaks Feature Set& Only the bursty behavior of the browsers. Can be useful in the case a smart user (adversarial opponent) changes all the protocols parameters.
 \\\hline
			New Feature Set& A new set of features (presented in Table \ref{tab:newfeatures}), based on a comprehensive network traffic analysis, in which we
identified traffic parameters that differentiate between different operating systems and browsers. The set of features included SSL features, TCP features and the bursty behavior of the browsers (peaks) which is defined as a section of traffic preceded and followed by silence. Note, we define it as new in order to be able to split later the there groups (SSL features, TCP features and peak features). 
\\\hline
			Common Stats Feature Set&  Only statistics parameters from Table \ref{tab:basefeatures} 
\\\hline
			Statistics & Only statistics parameters from Tables \ref{tab:basefeatures}, \ref{tab:newfeatures}. Can be useful in the case a smart user changes all the protocols parameters and we can only use the parameters related to the arriving packets.
 \\\hline
			Combined Feature Set& All available features combined (all the sets above). 
	\\\hline
			Combined no peaks Feature Set& All available features combined minus the peak features. Can be useful in the case a smart user changes the traffic pattern of the session.
 \\\hline
			Combined no SSL Feature Set& All available features combined minus the SSL related features. Can be useful in the case a smart user changes the SSL parameters (in the header, e.g. Cipher Suite). 
 \\\hline
			Combined no TCP Feature Set& All available features combined minus the TCP protocol related features. Can be useful in the case a smart user changes the TCP parameters (in the header, e.g. MSS size). 
	\\\hline
 \end{tabular}
  \caption{Feature Sets}
  \label{tab:FS}
\end{table}

\begin{figure}
	\centering
	\trafficTracesPlota{chrome_costarica4k_720_auto.tex}
	\caption{An example of bursty behavior of browser traffic. }
	\label{fig:chrome_720_auto_costarica_wireshark}
\end{figure}

\input{acc_fig_new_short.tex}

\input{confusion_fig.tex}

\input{acc_fig_new.tex}

\input{vpn_chiper_short.tex}


\input{var_training_data_set_size.tex}

\input{acc_realtime.tex}

\input{acc_runtime_split.tex}

\subsection{Learning}
\label{sec:Learning}
In this section we describe our machine learning methodology. Supervised classification learning methods learn a
classification function from a set of pre-labeled examples. The classification function is then used for
classifying unseen test examples. There are two types of supervised classification learning methods: lazy and eager. Lazy learning algorithms store the training data as is and then apply a classification function on a new test example where the classification function is parameterized with the pre-labeled training examples. Eager learning algorithms, on the other hand, carry out a learning process on the pre-labeled training examples. Eager learning algorithms often perform better since the offline learning stage increases robustness to noise.

The lazy machine learning algorithm we selected is the nearest neighbor algorithm \cite{cover1967nearest}. In this algorithm, the classification function computes similarities between a new test sample and all pre-labeled examples. The test sample is then assigned to the class of the most similar example from the training data.

The first eager machine learning algorithm we chose was the Support Vector Machine (SVM) \cite{SVM}. The binary linear SVM models training
examples as points in space, and then divides the space using a hyperplane to give the best separation between the two classes. We used
the LIBSVM package \cite{CC01a} which implements the one-vs-one multiclass scheme. We used three versions of the SVM:

\begin{itemize}
	\item SVM+RBF - SVM with the Radial Basis Function (RBF) as the kernel function.
	\item SVM+SIM - SVM with threshold similarities to the training samples as features \cite{chen2009similarity}. Similarity is bounded by a threshold. This is a heuristic based on the assumption that dissimilar samples add noise rather than information. The threshold similarity value and the distance function are chosen by the cross-validation process.
		\[
			u,v \in samples, 1- \frac{min(distance(u,v),threshold)}{threshold}
	\]
	\item SVM+MAP - SVM with modified RBF similarities as features \cite{chen2009similarity} where the similarity was the same as in the RBF function except for the distance function which was not necessarily squared Euclidean. The distance function and gamma values are chosen by cross-validation process.
\[
			u,v \in samples, exp(-\gamma \cdot distance(u,v))
	\]
\end{itemize}

The second eager machine learning algorithm we chose was the Random
Forest algorithm \cite{breiman2001random,ho1998random}. The Random Forest 
algorithm grows multiple trees by randomly selecting subsets of features.
That is, trees are constructed in random subspaces \cite{breiman2001random,ho1998random}.

Thus in total we had five machine learning algorithms. We fine-tuned the
hyper-parameters of our machine learning algorithms through a grid
search procedure combined with 5-fold cross validation over the
training set. For all the machine learning algorithms, features were
scaled between zero and one at training and the same scaling factors
were used for the test set. A one-tailed t-test with a p-value $0.05$ for every pair of algorithm results vector was conducted to test whether the result of one algorithm was significantly better than the others. 

For the KNN algorithm, we used cross validation to choose the
number of neighbors over the set $\{4, 6, \ldots, 20\}$, the uniform or
distance-based weights, and the distance measures: Euclidean,
Manhattan, Chebyshev, Hamming and Canberra.  For the SVM algorithm, we
used cross validation to choose both the regularization parameter
of SVM, $C$, over the set $\{2^{-5}, 2^{-3}, \ldots, 2^{15}\}$ and for
the gamma parameter of RBF, over the set
$\{2^{-15},2^{-13},\ldots,2^{3}\}$. We used LIBSVM \cite{CC01a} to
train and test our dataset.  For the Random Forest algorithm we used
cross validation to choose the number of trees over the set $\{20,
40, \ldots, 120\}$. The machine learning algorithm code can be found in \cite{OurMLCode}.

\section{Results}
\label{Results}
We trained and tested on 70\% train and 30\% test splits five times; accuracy is reported as the average of these experiments (average error bar is $0.01\%$). We first show that the accuracy of the tuple classification using our combined set is higher compared to the baseline feature set (common set). Then we present our main innovation; that even a smart user that is aware of our machine learning system and tries to manipulate the system by changing the protocols parameters or using other tools has a negligible effect on the accuracy of our system. Then, we demonstrate the robustness of the system against a smart user, in the case where we cannot sniff for a lengthy period of time the input of the training set or when taking training samples limited by time (the first $x$ seconds/minutes of the sessions). To assess whether our system can run on real time system, we show that the training and testing time is low (test time in milliseconds).

The accuracy for the tuple \textless OS, Browser,
Application\textgreater{} classification with our feature set as compared to the baseline feature set (common set) is
presented in Figure \ref{fig:results_short}. The figure shows that the tuple \textless OS, Browser,
Application\textgreater{} classification of encrypted classification
is possible with high accuracy. In all the experiments, using our
combined set achieved the best results whereas the Random Forest
algorithm and SVM+MAP both achieved the highest accuracy. For tuple classification, the addition of our \newset{} increased the accuracy from
$93.52\%$ to $96.06\%$. A confusion matrix for the tuple accuracy is
shown in Fig \ref{fig:confusionmat}. It shows that the
classification is almost perfect, and most of the mistakes could be ascribed to unidentified labels which may have been be a correct answer that we
could not verify.

The accuracy for the tuple \textless OS, Browser, Application\textgreater{} classification with the nine feature sets is
presented in Figure \ref{fig:results}. Note that some of the feature sets (e.g. combined without TCP) are equivalent to the case where adversarial opponents change the protocols parameters. Using our \combinedset{} achieved the best results whereas the Random Forest algorithm and SVM+MAP both achieved the highest accuracy. When using a subset of the features to mitigate the influence of adversarial opponent also achieved high accuracy (between $94\%-95.8\%$).  

The influence of using network tools by the user (e.g. VPN) to affect our machine learning system can be seen in Figure \ref{fig:VPN_OS_Browser_short}. Although the opponent aggregates all the sessions together, our system is still able to classify the operation system and the browser with good accuracy ($81\%$). When classifying both the OS and the browser, the best performance is achieved using the SVM+MAP algorithm with combined feature set. Most other combinations exhibited relatively low performance.

The impact on accuracy when the user changes protocol parameters such as Cipher suites (SSL protocol parameter) can be seen in Fig. \ref{fig:Cipher_OS_Browser_short}. Although the opponent changes the number and the value of the cipher suites in the SSL header, our system is still able to classify the operation system and the browser with satisfactory  accuracy ($91\%$). When classifying both the OS and the Browser, the best performance was achieved using the RF algorithm with combined features set. Note that we decided to present changes in the Cipher suites and not in other cases since a number of works such as \cite{Husk2016} have used the Cipher suite as important parameters in their classification system. Note that, for the case of VPN and Cipher suites we have used the same lab architecture as despite above. We run over several operating systems and browsers with a VPN tool \cite{VPN} while running VPN samples and for changing the Cihper suites we used selenium. Overall we have a data set of more than 2000 samples.

To determine the effect of the training data set size, we ran an experiment with various training set sizes (between 50 samples and a full data-set, 14,443 =  0.7$\times$ 20,633 samples). Figure \ref{fig:var_data_size} shows that although the training set had fewer samples, the system still achieved a reasonable accuracy of 80\% for the baseline features and 85\% when the new features were added. We then investigated whether our system would achieve high accuracy in the case of short session time which means fewer data in each session. To do so, in the next experiments we built a training and test set from our session using up to $X$ seconds/minutes ($1$ second, $10$ seconds, $1$ minutes, $10$ minutes). Figure \ref{fig:short_sessions} shows that the accuracy decreased but was still high (close to $94\%$). Moreover, the effect of shortening the session slightly decrease after $1$ minute of session time, where the results for $1$ minute are comparable to the accuracy of $1$ second.

After observing that using short sessions did not have a pronounced effect on the results, we investigated the run time of training and testing. Figure \ref{fig:runningTimeTrainTest} shows that the training time of our algorithms was between $10$ seconds (RF) and $300$ seconds (SVM) whereas the run time of the testing was less than $1$ second, indicating that our testing algorithms can run on real time networks.

\section{Possible Countermeasures and Limitations}
\label{Possible Countermeasures and Limitations}
Although users and service providers often assume that if they use the right encryption and authentication mechanisms their communications are secure, they are still vulnerable. As presented in this paper, it is possible to develop classifiers for TLS/SSL encrypted traffic that are able to discriminate between OSs, browsers and applications. We showed that changing protocol parameters such as cipher suites or using VPNs as counter-measures at test time reduced accuracy, but our system was able to identify the information with reasonable accuracy. While it is beyond the scope of the paper to investigate all possible countermeasures, we discuss some related issues.

Padding techniques are another simple countermeasure which may be effective against traffic analysis approaches. However, padding countermeasures are already standardized in TLS, explicitly to “frustrate attacks on a protocol that are based on analysis of the lengths of exchanged messages” \cite{TLS}.The intuition is that the information is not hidden efficiently, and the analysis of these features may still allow analysis. 

The main limitation of our approach has to do with our implementation of supervised learning algorithms. This technique is generally more efficient than unsupervised learning since it benefits from knowing each class of interest. However, it has three main drawbacks: (1) The training dataset has to be labeled by a human expert; (2) It is hard to recognize classes of events that have not been used during the training phase; (3) Upgrades (OS, browsers, applications) can change traffic patterns.

We mitigated the first limitation by using an automatic approach to label the network traces collected for the training phase. However, the second limitation cannot be addressed without revising the entire approach. In order to mitigate the third limitation we connected our lab to a VPN network that added another layer of encryption and we changed the Cipher suites number and values of the browsers. In both cases we used the same classifiers when classifying the operating system and the browser. Fig. \ref{fig:VPN_Cipher_short} shows that in both cases although the results were slightly affected, the accuracy remained high.

\section{Conclusion and Future Work}
\label{Conclusions and Future Work}
The framework proposed in this paper is able to classify encrypted network traffic and infer which operating system, browser and application the user is employing even in the case where a user tries to manipulate the traffic. We also showed that despite the use of SSL/TLS, our traffic analysis approach is an effective tool. An eavesdropper can easily leverage the information about the user to fit an optimal attack vector. An interesting extension of this work would be to add more user abilities and develop a system that will be robust to other types of attacks such as fake samples.

\section*{Acknowledgment}
This research was supported by the InfoMedia consortium.

\bibliographystyle{unsrt}
\bibliography{DSC_IEEE}

\end{document}

%% file: macros.tex
\newcommand{\baseset}{common features}
\newcommand{\newset}{new features}
\newcommand{\combinedset}{base + new features}

\newcommand{\trafficTracesPlota}[1]{
    \begin{tikzpicture}[scale=0.7]
      \begin{axis}[
          scale=1,
          xlabel={Time(sec)},
          ylabel={$\frac{Bytes}{sec}$},
          ymin=-300000,
          ymax=3000000,
          xmin=-10,
          xmax=250,
        ]
        \addplot[blue] table [col sep=comma]{#1};
      \end{axis}
    \end{tikzpicture}
}

\newcommand{\accResults}[2]
{
	\pgfplotsset
	 {compat=1.11,
    /pgfplots/ybar legend/.style=
		{
       /pgfplots/legend image code/.code={%
       \draw[##1,/tikz/.cd,yshift=-0.25em]
        (0cm,0cm) rectangle (3pt,0.8em);},
     },
   }
	\begin{tikzpicture}
  \pgfplotsset{compat=1.8}
	\begin{axis}[
					ybar,
					bar width=1pt,
					height = 7cm,
					width = 8cm,
					every axis plot/.append style={fill},
					font=\scriptsize,
          ytick={80,90,95,100},%
					yticklabels={80,90,95,100},%
					xticklabels={,KNN,SVM+RBF,RF,SVM+MAP,SVM+SIM},%
          xlabel={Machine Learning Algorithm},
          ylabel={Accuracy [\%]},
		      ymax=#2, ymin = 70,
			  grid,
              legend cell align={left},
					]
      \addplot [blue] table[x=ML, y=Peaks, col sep=comma]{#1};
      \addplot [red] table[x=ML, y=Common_stats, col sep=comma]{#1};
      \addplot [orange] table[x=ML, y=Stats, col sep=comma]{#1}; 
      \addplot [gray] table[x=ML, y=Common, col sep=comma]{#1};
	    \addplot [black] table[x=ML, y=New, col sep=comma]{#1};
      \addplot [green] table[x=ML, y=Combined_no_SSL, col sep=comma]{#1};
	    \addplot [lime] table[x=ML, y=Combined_no_peaks, col sep=comma]{#1};
      \addplot [cyan] table[x=ML, y=Combined_no_TCP, col sep=comma]{#1};
	    \addplot [yellow] table[x=ML, y=Combined_Features, col sep=comma]{#1};
    \end{axis}
    \end{tikzpicture}
}

\newcommand{\accResultsFirst}[2]
{
	\pgfplotsset
	 {compat=1.11,
    /pgfplots/ybar legend/.style=
		{
       /pgfplots/legend image code/.code={%
       \draw[##1,/tikz/.cd,yshift=-0.25em]
        (0cm,0cm) rectangle (3pt,0.8em);},
     },
   }
	\begin{tikzpicture}
  \pgfplotsset{compat=1.8}
	\begin{axis}[
					legend to name=barLagendName,
					ybar,
					bar width=1pt,
					height = 7cm,
					width = 8cm,
					every axis plot/.append style={fill},
					font=\scriptsize,
          ytick={80,90,95,100},%
					yticklabels={80,90,95,100},%
					xticklabels={,KNN,SVM+RBF,RF,SVM+MAP,SVM+SIM},%
          xlabel={Machine Learning Algorithm},
          ylabel={Accuracy [\%]},
		      ymax=#2, ymin = 70,
			  grid,
              legend cell align={left},
					]
      \addplot [blue] table[x=ML, y=Peaks, col sep=comma]{#1};
      \addplot [red] table[x=ML, y=Common_stats, col sep=comma]{#1};
      \addplot [orange] table[x=ML, y=Stats, col sep=comma]{#1}; 
      \addplot [gray] table[x=ML, y=Common, col sep=comma]{#1};
	    \addplot [black] table[x=ML, y=New, col sep=comma]{#1};
      \addplot [green] table[x=ML, y=Combined_no_SSL, col sep=comma]{#1};
	    \addplot [lime] table[x=ML, y=Combined_no_peaks, col sep=comma]{#1};
      \addplot [cyan] table[x=ML, y=Combined_no_TCP, col sep=comma]{#1};
	    \addplot [yellow] table[x=ML, y=Combined_Features, col sep=comma]{#1};
			
			\legend{Peaks,Common stats,Stats,Common,Only New,Combined (ours) no SSL,Combined (ours) no peaks,Combined (ours) no TCP,Combined (ours)}
    \end{axis}
    \end{tikzpicture}

}

\newcommand{\accResultsShort}[2]
{
	\pgfplotsset
	 {compat=1.11,
    /pgfplots/ybar legend/.style=
		{
       /pgfplots/legend image code/.code={%
       \draw[##1,/tikz/.cd,yshift=-0.25em]
        (0cm,0cm) rectangle (3pt,0.8em);},
     },
   }
	\begin{tikzpicture}
  \pgfplotsset{compat=1.8}
	\begin{axis}[
					ybar,
					bar width=1pt,
					height = 7cm,
					width = 8cm,
					every axis plot/.append style={fill},
					font=\scriptsize,
          ytick={90,92,95,97,100},%
					yticklabels={90,92,95,97,100},%
					xticklabels={,KNN,SVM+RBF,RF,SVM+MAP,SVM+SIM},%
          xlabel={Machine Learning Algorithm},
          ylabel={Accuracy [\%]},
		      ymax=#2, ymin = 85,
			  grid,
              legend cell align={left},
					]
      \addplot [gray] table[x=ML, y=Common, col sep=comma]{#1};
	    \addplot [yellow] table[x=ML, y=Combined_Features, col sep=comma]{#1};
    \end{axis}
    \end{tikzpicture}
}

\newcommand{\accResultsFirstShort}[2]
{
	\pgfplotsset
	 {compat=1.11,
    /pgfplots/ybar legend/.style=
		{
       /pgfplots/legend image code/.code={%
       \draw[##1,/tikz/.cd,yshift=-0.25em]
        (0cm,0cm) rectangle (3pt,0.8em);},
     },
   }
	\begin{tikzpicture}
  \pgfplotsset{compat=1.8}
	\begin{axis}[
					legend to name=barLagendName,
					ybar,
					bar width=1pt,
					height = 7cm,
					width = 8cm,
					every axis plot/.append style={fill},
					font=\scriptsize,
          ytick={90,92,95,97,100},%
					yticklabels={90,92,95,97,100},%
					xticklabels={,KNN,SVM+RBF,RF,SVM+MAP,SVM+SIM},%
          xlabel={Machine Learning Algorithm},
          ylabel={Accuracy [\%]},
		      ymax=#2, ymin = 85,
			  grid,
              legend cell align={left},
					]
      \addplot [red] table[x=ML, y=Common, col sep=comma]{#1};
	    \addplot [green] table[x=ML, y=Combined_Features, col sep=comma]{#1};
			
			\legend{Common,Combined (ours)}
    \end{axis}
    \end{tikzpicture}

}

\newcommand{\VPNCipherFirst}[2]
{
	\pgfplotsset
	 {compat=1.11,
    /pgfplots/ybar legend/.style=
		{
       /pgfplots/legend image code/.code={%
       \draw[##1,/tikz/.cd,yshift=-0.25em]
        (0cm,0cm) rectangle (3pt,0.8em);},
     },
   }
	\begin{tikzpicture}
  \pgfplotsset{compat=1.8}
	\begin{axis}[
					legend to name=barLagendNameVPN,
					ybar,
					bar width=1pt,
					height = 7cm,
					width = 8cm,
					every axis plot/.append style={fill},
					font=\scriptsize,
          ytick={0, 10, 20, 30, 40, 50, 60, 70, 80,90,95,100},%
					yticklabels={0, 10, 20, 30, 40, 50, 60, 70, 80,90,95,100},%
					xticklabels={,KNN,SVM+RBF,RF,SVM+MAP,SVM+SIM},%
          xlabel={Machine Learning Algorithm},
          ylabel={Accuracy [\%]},
		      ymax=#2, ymin = 0,
					grid,
					legend cell align={left},
					]
      \addplot [blue] table[x=ML, y=Peaks, col sep=comma]{#1};
      \addplot [red] table[x=ML, y=Common_stats, col sep=comma]{#1};
      \addplot [orange] table[x=ML, y=Stats, col sep=comma]{#1}; 
      \addplot [gray] table[x=ML, y=Common, col sep=comma]{#1};
	    \addplot [black] table[x=ML, y=New, col sep=comma]{#1};
      \addplot [green] table[x=ML, y=Combined_no_SSL, col sep=comma]{#1};
	    \addplot [lime] table[x=ML, y=Combined_no_peaks, col sep=comma]{#1};
      \addplot [cyan] table[x=ML, y=Combined_no_TCP, col sep=comma]{#1};
	    \addplot [yellow] table[x=ML, y=Combined_Features, col sep=comma]{#1};
			
			\legend{Peaks,Common stats,Stats,Common,Only New,Combined (ours) no SSL,Combined (ours) no peaks,Combined (ours) no TCP,Combined (ours)}
    \end{axis}
    \end{tikzpicture}

}

\newcommand{\VPNCipher}[2]
{
	\pgfplotsset
	 {compat=1.11,
    /pgfplots/ybar legend/.style=
		{
       /pgfplots/legend image code/.code={%
       \draw[##1,/tikz/.cd,yshift=-0.25em]
        (0cm,0cm) rectangle (3pt,0.8em);},
     },
   }
	\begin{tikzpicture}
  \pgfplotsset{compat=1.8}
	\begin{axis}[
					legend to name=barLagendName,
					ybar,
					bar width=1pt,
					height = 7cm,
					width = 8cm,
					every axis plot/.append style={fill},
					font=\scriptsize,
          ytick={0, 10, 20, 30, 40, 50, 60, 70, 80,90,95,100},%
					yticklabels={0, 10, 20, 30, 40, 50, 60, 70, 80,90,95,100},%
					xticklabels={,KNN,SVM+RBF,RF,SVM+MAP,SVM+SIM},%
          xlabel={Machine Learning Algorithm},
          ylabel={Accuracy [\%]},
		      ymax=#2, ymin = 0,
					grid,
					legend cell align={left},
					]
      \addplot [blue] table[x=ML, y=Peaks, col sep=comma]{#1};
      \addplot [red] table[x=ML, y=Common_stats, col sep=comma]{#1};
      \addplot [orange] table[x=ML, y=Stats, col sep=comma]{#1}; 
      \addplot [gray] table[x=ML, y=Common, col sep=comma]{#1};
	    \addplot [black] table[x=ML, y=New, col sep=comma]{#1};
      \addplot [green] table[x=ML, y=Combined_no_SSL, col sep=comma]{#1};
	    \addplot [lime] table[x=ML, y=Combined_no_peaks, col sep=comma]{#1};
      \addplot [cyan] table[x=ML, y=Combined_no_TCP, col sep=comma]{#1};
	    \addplot [yellow] table[x=ML, y=Combined_Features, col sep=comma]{#1};
			
			\legend{Peaks,Common stats,Stats,Common,Only New,Combined (ours) no SSL,Combined (ours) no peaks,Combined (ours) no TCP,Combined (ours)}
    \end{axis}
    \end{tikzpicture}

}

\newcommand{\TrainTestFirst}[2]
{
	\begin{tikzpicture}
					 \begin{loglogaxis}
					[
					 legend to name=TimeLagendName2,
						grid,
						height = 6cm,
						width = 6cm,
						font=\scriptsize,
						xtick = {50,100,200,500,1000,5000,14443},
						xticklabels={50,100,200,500,1000,5000,14443},%
						xlabel={Training Dataset Size},
						ylabel={Time [sec]},
						ymax=#2, ymin = 0,
						log basis x={2},
            legend cell align={left},
                    ]	
					\addplot [blue,dashdotted, mark=x,mark options={scale=1,solid}] table[x=Training_Data_Set_Size, y=Peaks, col sep=comma]{#1};
					\addplot [red,loosely dotted,mark=+,mark options={scale=1,solid}] table[x=Training_Data_Set_Size, y=Common_stats, col sep=comma]{#1};
				  \addplot [orange,loosely dotted,mark=+,mark options={scale=1,solid}] table[x=Training_Data_Set_Size, y=Stats, col sep=comma]{#1}; 
					\addplot [gray, mark=square*, mark options={scale=1,solid}] table[x=Training_Data_Set_Size, y=Common, col sep=comma]{#1};
					\addplot [black, mark=square*, mark options={scale=1,solid}] table[x=Training_Data_Set_Size, y=New, col sep=comma]{#1};
				  \addplot [green,dashed, mark=o,mark options={scale=1,solid}] table[x=Training_Data_Set_Size, y=Combined_no_SSL, col sep=comma]{#1};
				  \addplot [lime,dashed, mark=o,mark options={scale=1,solid}] table[x=Training_Data_Set_Size, y=Combined_no_peaks, col sep=comma]{#1};           
				  \addplot [cyan,dashdotted, mark=x,mark options={scale=1,solid}] table[x=Training_Data_Set_Size, y=Combined_no_TCP, col sep=comma]{#1};           
					\addplot [yellow,loosely dotted,mark=+,mark options={scale=1,solid}] table[x=Training_Data_Set_Size, y=Combined_Features, col sep=comma]{#1};
					\legend{Peaks,Common stats,Stats,Common,Only New,Combined (ours) no SSL,Combined (ours) no peaks,Combined (ours) no TCP,Combined (ours)}
					\end{loglogaxis}
	\end{tikzpicture}
}

\newcommand{\TrainTest}[2]
{
	\pgfplotsset
	 {compat=1.11,
    /pgfplots/ybar legend/.style=
		{
       /pgfplots/legend image code/.code={%
       \draw[##1,/tikz/.cd,yshift=-0.25em]
        (0cm,0cm) rectangle (3pt,0.8em);},
     },
   }
	\begin{tikzpicture}
					 \begin{axis}
					  [
						grid,
						height = 6cm,
						width = 6cm,
						font=\scriptsize,
						xtick = {50,5000,14443},
						xticklabels={50,5000,14443},%
						xlabel={Training Dataset Size},
						ylabel={Time [sec]},
						ymax=#2, ymin = 0,
					]	
					\addplot [blue,dashdotted, mark=x,mark options={scale=1,solid}] table[x=Training_Data_Set_Size, y=Peaks, col sep=comma]{#1};
					\addplot [red,loosely dotted,mark=+,mark options={scale=1,solid}] table[x=Training_Data_Set_Size, y=Common_stats, col sep=comma]{#1};
				  \addplot [orange,loosely dotted,mark=+,mark options={scale=1,solid}] table[x=Training_Data_Set_Size, y=Stats, col sep=comma]{#1}; 
					\addplot [gray, mark=square*, mark options={scale=1,solid}] table[x=Training_Data_Set_Size, y=Common, col sep=comma]{#1};
					\addplot [black, mark=square*, mark options={scale=1,solid}] table[x=Training_Data_Set_Size, y=New, col sep=comma]{#1};
				  \addplot [green,dashed, mark=o,mark options={scale=1,solid}] table[x=Training_Data_Set_Size, y=Combined_no_SSL, col sep=comma]{#1};
				  \addplot [lime,dashed, mark=o,mark options={scale=1,solid}] table[x=Training_Data_Set_Size, y=Combined_no_peaks, col sep=comma]{#1};           
				  \addplot [cyan,dashdotted, mark=x,mark options={scale=1,solid}] table[x=Training_Data_Set_Size, y=Combined_no_TCP, col sep=comma]{#1};           
					\addplot [yellow,loosely dotted,mark=+,mark options={scale=1,solid}] table[x=Training_Data_Set_Size, y=Combined_Features, col sep=comma]{#1};
					\end{axis}
	\end{tikzpicture}
}

\newcommand{\TrainTestLog}[2]
{
	\begin{tikzpicture}
					 \begin{loglogaxis}
					[
						grid,
						height = 6cm,
						width = 6cm,
						font=\scriptsize,
						xtick = {50,100,200,500,1000,5000,14443},
						xticklabels={50,100,200,500,1000,5000,14443},%
						xlabel={Training Dataset Size},
						ylabel={Time [sec]},
						ymax=#2, ymin = 0,
						log basis x={2},
                        legend cell align={left},
					]	
					\addplot [blue,dashdotted, mark=x,mark options={scale=1,solid}] table[x=Training_Data_Set_Size, y=Peaks, col sep=comma]{#1};
					\addplot [red,loosely dotted,mark=+,mark options={scale=1,solid}] table[x=Training_Data_Set_Size, y=Common_stats, col sep=comma]{#1};
				  \addplot [orange,loosely dotted,mark=+,mark options={scale=1,solid}] table[x=Training_Data_Set_Size, y=Stats, col sep=comma]{#1}; 
					\addplot [gray, mark=square*, mark options={scale=1,solid}] table[x=Training_Data_Set_Size, y=Common, col sep=comma]{#1};
					\addplot [black, mark=square*, mark options={scale=1,solid}] table[x=Training_Data_Set_Size, y=New, col sep=comma]{#1};
				  \addplot [green,dashed, mark=o,mark options={scale=1,solid}] table[x=Training_Data_Set_Size, y=Combined_no_SSL, col sep=comma]{#1};
				  \addplot [lime,dashed, mark=o,mark options={scale=1,solid}] table[x=Training_Data_Set_Size, y=Combined_no_peaks, col sep=comma]{#1};           
				  \addplot [cyan,dashdotted, mark=x,mark options={scale=1,solid}] table[x=Training_Data_Set_Size, y=Combined_no_TCP, col sep=comma]{#1};           
					\addplot [yellow,loosely dotted,mark=+,mark options={scale=1,solid}] table[x=Training_Data_Set_Size, y=Combined_Features, col sep=comma]{#1};
					\end{loglogaxis}
	\end{tikzpicture}
}

\newcommand{\varsTrainSizes}[1]
{
  \begin{tikzpicture}
	\begin{axis}[
		grid,
		height = 7cm,
		width = 5.5cm,
		font=\scriptsize,
		xtick = {50,100,200,500,1000,5000,14443},
		ytick={0,10,20,30,40,50,60,70,80,90,100},%
		yticklabels={0,10,20,30,40,50,60,70,80,90,100},%
		xticklabels={50,100,200,500,1000,5000,14443},%
		legend style={font=\scriptsize,at={(axis cs:0,-10)},anchor=south west},
		xlabel={Training Dataset Size},
		ylabel={Accuracy [\%]},
		ymax=102, ymin = 40,
		xmode=log,
		log basis x={2},
		legend cell align={left},
	  ]	
	  \addplot [blue,dashdotted, mark=x,mark options={scale=1,solid}] table[x=Training_Data_Set_Size, y=Peaks, col sep=comma]{#1};
	  \addplot [red,loosely dotted,mark=+,mark options={scale=1,solid}] table[x=Training_Data_Set_Size, y=Common_stats, col sep=comma]{#1};
	  \addplot [orange,loosely dotted,mark=+,mark options={scale=1,solid}] table[x=Training_Data_Set_Size, y=Stats, col sep=comma]{#1}; 
	  \addplot [gray, mark=square*, mark options={scale=1,solid}] table[x=Training_Data_Set_Size, y=Common, col sep=comma]{#1};
	  \addplot [black, mark=square*, mark options={scale=1,solid}] table[x=Training_Data_Set_Size, y=New, col sep=comma]{#1};
	  \addplot [green,dashed, mark=o,mark options={scale=1,solid}] table[x=Training_Data_Set_Size, y=Combined_no_SSL, col sep=comma]{#1};
	  \addplot [lime,dashed, mark=o,mark options={scale=1,solid}] table[x=Training_Data_Set_Size, y=Combined_no_peaks, col sep=comma]{#1};           
	  \addplot [cyan,dashdotted, mark=x,mark options={scale=1,solid}] table[x=Training_Data_Set_Size, y=Combined_no_TCP, col sep=comma]{#1};           
	  \addplot [yellow,loosely dotted,mark=+,mark options={scale=1,solid}] table[x=Training_Data_Set_Size, y=Combined_Features, col sep=comma]{#1};
	\end{axis}
  \end{tikzpicture}
}

\newcommand{\varsTrainSizesFirst}[1]
{
	         \begin{tikzpicture}
			   \begin{axis}
				 [
                   grid,
				   height = 7cm,
				   width = 5.5cm,
				   font=\scriptsize,
				   xtick = {50,100,200,500,1000,5000,14443},
				   ytick={0,10,20,30,40,50,60,70,80,90,100},%
				   yticklabels={0,10,20,30,40,50,60,70,80,90,100},%
				   xticklabels={50,100,200,500,1000,5000,14443},%
				   legend style={font=\scriptsize,at={(axis cs:0,-10)},anchor=south west},
				   xlabel={Training Dataset Size},
				   ylabel={Accuracy [\%]},
				   ymax=102, ymin = 40,
				   xmode=log,
				   log basis x={2},
				   legend cell align={left},
                   legend to name=p,
				 ]	
				 \addplot [blue,dashdotted, mark=x,mark options={scale=1,solid}] table[x=Training_Data_Set_Size, y=Peaks, col sep=comma]{#1};
				 \addplot [red,loosely dotted,mark=+,mark options={scale=1,solid}] table[x=Training_Data_Set_Size, y=Common_stats, col sep=comma]{#1};
				 \addplot [orange,loosely dotted,mark=+,mark options={scale=1,solid}] table[x=Training_Data_Set_Size, y=Stats, col sep=comma]{#1}; 
				 \addplot [gray, mark=square*, mark options={scale=1,solid}] table[x=Training_Data_Set_Size, y=Common, col sep=comma]{#1};
				 \addplot [black, mark=square*, mark options={scale=1,solid}] table[x=Training_Data_Set_Size, y=New, col sep=comma]{#1};
				 \addplot [green,dashed, mark=o,mark options={scale=1,solid}] table[x=Training_Data_Set_Size, y=Combined_no_SSL, col sep=comma]{#1};
				 \addplot [lime,dashed, mark=o,mark options={scale=1,solid}] table[x=Training_Data_Set_Size, y=Combined_no_peaks, col sep=comma]{#1};           
				 \addplot [cyan,dashdotted, mark=x,mark options={scale=1,solid}] table[x=Training_Data_Set_Size, y=Combined_no_TCP, col sep=comma]{#1};           
				 \addplot [yellow,loosely dotted,mark=+,mark options={scale=1,solid}] table[x=Training_Data_Set_Size, y=Combined_Features, col sep=comma]{#1};
                 \legend{Peaks,Common stats,Stats,Common,Only New,Combined (ours) no SSL,Combined (ours) no peaks,Combined (ours) no TCP,Combined (ours)}
			   \end{axis}
	         \end{tikzpicture}
}

\newcommand{\mycomment}[1]{}

\newcommand{\etal}{et al.\@ }

\newcommand{\naive}{na\"{\i}ve}


%% file: stat_fig.tex
\pgfplotstableread[col sep=comma,header=false]{
  OSX	Chrome Unidentified,0.02 
  OSX	Chrome	Google-Background,0.09 
  Windows	IExplorer	Google-Background,0.13 		
  Ubuntu	Non-Browser	Microsoft-Background,0.13 
  Windows	IExplorer	Unidentified,0.13  
  OSX	Chrome	Twitter,0.27 
  Ubuntu	Firefox	Facebook,0.40 
  Windows	Firefox	Unidentified,0.44
  Ubuntu	Chrome	Facebook,0.46
  Windows	Chrome	Unidentified,0.64
  Windows Non-Browser Dropbox,0.67
  Ubuntu	Chrome	Youtube,0.67 
  Windows	Non-Browser	Teamviewer,0.71 		
  Ubuntu	Firefox	Youtube,0.74 
  OSX	Safari	Twitter,1.20 
  Windows	Firefox	Google-Background,1.73 
  Ubuntu	Chrome	Twitter,1.75 
  Ubuntu	Chrome	Google-Background,1.90 
  Ubuntu	Firefox	Unidentified,1.95 
  OSX	Safari	Unidentified,2.34 
  Ubuntu	Firefox	Twitter,2.72 
  Windows	Chrome	Google-Background,2.82 
  Ubuntu	Chrome	Unidentified,3.42 		
  OSX	Safari	Youtube,4.03 
  OSX	Safari	Google-Background,4.94 
  Windows	Firefox	Twitter,5.25 
  Windows	Chrome	Twitter,7.02 
  Windows	Non-Browser	Microsoft-Background,7.81 
  Ubuntu	Firefox	Google-Background,13.13 
  Windows	IExplorer	Twitter,32.34 
}\tupledata

\pgfplotstableread[col sep=comma,header=false]{
  Windows,59.73
  OSX,12.93
  Ubuntu,27.33
}\osdata

\pgfplotstableread[col sep=comma,header=false]{
  Non-Browser,9.33
  Safari,12.53
  Chrome,19.11
  Firefox,26.40
  IExplorer,32.60
}\browserdata

\pgfplotstableread[col sep=comma,header=false]{
  Dropbox,0.6736780885
  Teamviewer,0.7124509281
  Facebook,0.8675422866
  Youtube,5.457277177
  Microsoft-Background,7.948432123
  Unidentified,8.966219163
  Google-Background,24.77584452
  Twitter,50.59855571
}\appdata

\pgfplotsset{
  percentage plot/.style={
    point meta=explicit,
    xticklabel=\pgfmathprintnumber{\tick}\,$\%$,
    enlarge x limits={upper,value=0},
    visualization depends on={x \as \originalvalue}
  }
}

\begin{figure}[htb]
\begin{minipage}[l]{0.95\linewidth}
\begin{tikzpicture}
\begin{axis}[
	unit vector ratio*=1 4 1,
	font=\scriptsize,
	xmajorgrids=true,
			xmin=0,
			xmax=35,
	axis on top,
	xtick = {0, 15, 30},
	height = 8 cm,
	xlabel=Percentage,
	percentage plot,
	xbar=0pt,
	bar width=0.1cm,
	enlarge y limits=0.03, 
	every axis y label/.append style={yshift=-20cm},
	symbolic y coords={OSX Chrome	Unidentified, OSX	Chrome	Google-Background, Windows	IExplorer	Google-Background, Ubuntu	Non-Browser	Microsoft-Background,Windows	IExplorer	Unidentified,OSX	Chrome	Twitter,Ubuntu	Firefox	Facebook,Windows	Firefox	Unidentified, Ubuntu	Chrome	Facebook, Windows	Chrome	Unidentified, Windows Non-Browser Dropbox, Ubuntu	Chrome	Youtube, Windows	Non-Browser	Teamviewer, Ubuntu	Firefox	Youtube, OSX	Safari	Twitter, Windows	Firefox	Google-Background, Ubuntu	Chrome	Twitter, Ubuntu	Chrome	Google-Background, Ubuntu	Firefox	Unidentified, OSX	Safari	Unidentified, Ubuntu	Firefox	Twitter, Windows	Chrome	Google-Background, Ubuntu	Chrome	Unidentified, OSX	Safari	Youtube, OSX	Safari	Google-Background, Windows	Firefox	Twitter, Windows	Chrome	Twitter, Windows	Non-Browser	Microsoft-Background, Ubuntu	Firefox	Google-Background, Windows	IExplorer	Twitter},
ytick=data
]
\addplot table [x index = 1, y index = 0] {\tupledata};
\end{axis}
\end{tikzpicture}
\subcaption{Labels (tuple) statistics}
\label{mp:tuplestat}
\end{minipage}

\vspace{20pt}

\begin{minipage}[b]{0.35\linewidth}
\begin{tikzpicture}
\begin{axis}[
	font=\scriptsize,
	xmajorgrids=true,
			xmin=0,
			xmax=65,
	axis on top,
	xtick = {0, 30, 65},
	width= 4 cm,
	height = 4 cm,
	xlabel=Percentage,
	percentage plot,
	xbar=0pt,
	bar width=0.1cm,
	enlarge y limits=0.03, 
	every axis y label/.append style={yshift=-20cm},
	symbolic y coords={OSX, Ubuntu, Windows},
ytick=data
]
\addplot table [x index = 1, y index = 0] {\osdata};
\end{axis}
\end{tikzpicture}
\subcaption{OS statistics}
\label{mp:osstat}
\end{minipage}
\hspace{45pt}
\begin{minipage}[b]{0.35\linewidth}
\begin{tikzpicture}
\begin{axis}[
font=\scriptsize,
	xmajorgrids=true,
			xmin=0,
			xmax=35,
	axis on top,
	xtick = {0, 15, 35},
	width= 4 cm,
	height = 4 cm,
	xlabel=Percentage,
	percentage plot,
	xbar=0pt,
	bar width=0.1cm,
	enlarge y limits=0.03, 
	every axis y label/.append style={yshift=-20cm},
symbolic y coords={Non-Browser, Safari, Chrome, Firefox, IExplorer},
ytick=data
]
\addplot table [x index = 1, y index = 0] {\browserdata};
\end{axis}
\end{tikzpicture}
\subcaption{Browser statistics}
\label{mp:browswerstat}
\end{minipage}

\vspace{20pt}

\begin{minipage}[b]{0.6\linewidth}
\begin{tikzpicture}
\begin{axis}[
font=\scriptsize,
	xmajorgrids=true,
			xmin=0,
			xmax=55,
	axis on top,
	xtick = {0, 35, 55},
	width= 4 cm,
	height = 4 cm,
	xlabel=Percentage,
	percentage plot,
	xbar=0pt,
	bar width=0.1cm,
	enlarge y limits=0.03, 
	every axis y label/.append style={yshift=-20cm},
symbolic y coords={Dropbox,Teamviewer,Facebook,Youtube,Microsoft-Background,Unidentified,Google-Background,Twitter},
ytick=data
]
\addplot table [x index = 1, y index = 0] {\appdata};
\end{axis}
\end{tikzpicture}
\subcaption{Application statistics}
\label{mp:appstat}
\end{minipage}
\vspace{0.1cm}
\caption{Dataset statistics}
\label{datasetstat}
\end{figure}

%% file: acc_fig_new_short.tex
\begin{figure*}[t!]
\begin{minipage}[t]{0.4\linewidth}
\centering
\accResultsFirstShort{Tuple_results.tex}{102}
\subcaption{Tuple Accuracy Results}
\label{mp:tupleresults_short}
\end{minipage}
\hspace{20pt}
\begin{minipage}[t]{0.4\linewidth}
\centering
\accResultsShort{results_browser_clean.tex}{102}
\vspace{-11pt}
\subcaption{Browser Accuracy Results}
\label{mp:browsers_short}
\end{minipage}

\vspace{10pt}
\begin{minipage}[t]{0.4\linewidth}
\centering
\accResultsShort{results_os_clean.tex}{102}
\vspace{-11pt}
\subcaption{OS Accuracy Results}
\label{mp:oss_short}
\end{minipage}
\hspace{20pt}
\begin{minipage}[t]{0.4\linewidth}
\centering
\accResultsShort{results_app_clean.tex}{102}
\vspace{-11pt}
\subcaption{Application Accuracy Results}
\label{mp:applications_short}
\end{minipage}

\begin{center}
\ref{barLagendName}
\end{center}
\caption{Accuracy results for KNN, SVM-RBF, SVM-MAP, SVM-SIM, RF. Adding our new features increased the accuracy to $96.6\%$. The captions are the same for the four subfigures}
\label{fig:results_short}
\end{figure*}

%% file: confusion_fig.tex
\pgfplotstableset{
  every head row/.style={
    before row=\hline,
    after row=\hline,
    typeset cell/.code={
      \ifnum\pgfplotstablecol=\pgfplotstablecols
      \pgfkeyssetvalue{/pgfplots/table/@cell content}{\rotatebox{90}{##1}\\}%
      \else
      \pgfkeyssetvalue{/pgfplots/table/@cell content}{\rotatebox{90}{##1}&}%
      \fi
    }
  },
  every last row/.style={
    after row=\hline
  },
  every last column/.style={
    column type/.add={}{|}
  },
  every first column/.style={
    column type=|l|
  },
  color cells/.style={
    col sep=&,
    row sep=\\,
    string type,
    postproc cell content/.code={%
      \pgfkeysalso{@cell content=\rule{0cm}{0.4ex}%
        \pgfmathsetmacro\y{max((##1 * 100), 0}%
        \edef\temp{\noexpand\cellcolor{black!\y}}\temp%
        \pgfmathtruncatemacro\x\y%
        \ifnum\x>50 \color{white}\fi%
        ##1}%
    },
    columns/x/.style={
        column name={},
        postproc cell content/.code={}
      }
  }
}

\setlength{\tabcolsep}{3pt}

\begin{figure*}[t!]

  \begin{minipage}[b]{0.2cm}
    \rotatebox{90}{\hspace{2cm} \tiny Real labels}
  \end{minipage}%
  \begin{minipage}[b]{.9\linewidth}
    \centering
    \begin{tabular}{c}
    \hspace{3cm} \tiny Predicted labels \\

    \end{tabular}
    \tiny
    \pgfplotstabletypeset[color cells]{
      x&Windows IExplorer Twitter&Ubuntu Firefox Google-Services&Windows Non-Browser Microsoft-Services&Windows Chrome Twitter&Windows Firefox Twitter&OSX Safari Google-Services&OSX Safari Youtube&Ubuntu Chrome Unidentified&Windows Chrome Google-Services&Ubuntu Firefox Twitter&OSX Safari Unidentified&Ubuntu Firefox Unidentified&Ubuntu Chrome Google-Services&Ubuntu Chrome Twitter&Windows Firefox Google-Services&OSX Safari Twitter&Ubuntu Firefox Youtube&Windows Non-Browser Teamviewer&Ubuntu Chrome Youtube&Windows Non-Browser Dropbox&Windows Chrome Unidentified&Ubuntu Chrome Facebook&Windows Firefox Unidentified&Ubuntu Firefox Facebook&OSX Chrome Twitter&Windows IExplorer Unidentified&Ubuntu Non-Browser Skype&Windows IExplorer Google-Services&OSX Chrome Google-Services&OSX Chrome Unidentified\\
      Windows IExplorer Twitter&1&0&0&0&0&0&0&0&0&0&0&0&0&0&0&0&0&0&0&0&0&0&0&0&0&0&0&0&0&0\\
      Ubuntu Firefox Google-Services&0&.97&0&0&0&0&0&0&0&0&0&0&.01&0&0&0&0&0&0&0&0&0&0&0&0&0&0&0&0&0\\
      Windows Non-Browser Microsoft-Services&0&0&.99&0&0&0&0&0&0&0&0&0&0&0&0&0&0&0&0&0&0&0&0&0&0&0&0&0&0&0\\
      Windows Chrome Twitter&0&0&0&.99&0&0&0&0&0&0&0&0&0&0&0&0&0&0&0&0&.01&0&0&0&0&0&0&0&0&0\\
      Windows Firefox Twitter&0&0&0&0&.98&0&0&0&0&0&0&0&0&0&0&0&0&0&0&0&0&0&.02&0&0&0&0&0&0&0\\
      OSX Safari Google-Services&0&0&0&0&0&.92&.04&0&0&0&.02&0&0&0&0&.02&0&0&0&0&0&0&0&0&0&0&0&0&0&0\\
      OSX Safari Youtube&0&0&0&0&0&.02&.97&.01&0&0&0&0&0&0&0&0&0&0&0&0&0&0&0&0&0&0&0&0&0&0\\
      Ubuntu Chrome Unidentified&0&0&0&0&0&0&0&.84&0&0&0&0&.07&.04&0&0&0&0&.01&0&0&.03&0&0&0&0&0&0&0&0\\
      Windows Chrome Google-Services&0&0&.01&.03&0&0&0&0&.94&0&0&0&0&0&.02&0&0&0&0&0&.01&0&0&0&0&0&0&0&0&0\\
      Ubuntu Firefox Twitter&0&0&0&0&0&0&0&0&0&.95&0&.03&0&0&0&0&.01&0&0&0&0&0&0&0&0&0&0&0&0&0\\
      OSX Safari Unidentified&0&0&0&0&0&.06&.01&0&0&0&.91&0&0&0&0&.01&0&0&0&0&0&0&0&0&0&0&0&0&0&0\\
      Ubuntu Firefox Unidentified&0&.02&0&0&0&0&0&0&0&.08&0&.87&0&0&0&0&.01&0&0&0&0&0&0&.03&0&0&0&0&0&0\\
      Ubuntu Chrome Google-Services&0&.07&0&0&0&0&0&.18&0&0&0&0&.73&0&0&0&0&0&.02&0&0&0&0&0&0&0&0&0&0&0\\
      Ubuntu Chrome Twitter&0&.02&0&0&0&0&0&.08&0&0&0&0&.03&.84&0&0&0&0&.01&0&0&.01&0&0&0&0&0&0&0&0\\
      Windows Firefox Google-Services&0&0&0&.01&0&0&0&0&.01&0&0&0&0&0&.97&0&0&0&0&0&0&0&.01&0&0&0&0&0&0&0\\
      OSX Safari Twitter&0&0&0&0&0&0&.06&0&0&0&.03&0&0&0&0&.91&0&0&0&0&0&0&0&0&0&0&0&0&0&0\\
      Ubuntu Firefox Youtube&0&.02&0&0&0&0&0&0&0&.02&0&.02&0&0&0&0&.93&0&0&0&0&0&0&0&0&0&0&0&0&0\\
      Windows Non-Browser Teamviewer&0&0&0&0&0&0&0&0&0&0&0&0&0&0&0&0&0&1&0&0&0&0&0&0&0&0&0&0&0&0\\
      Ubuntu Chrome Youtube&0&0&0&0&0&0&0&.07&0&0&0&0&.13&.04&0&0&0&0&.74&0&0&.02&0&0&0&0&0&0&0&0\\
      Windows Non-Browser Dropbox&0&0&0&0&0&0&0&0&0&0&0&0&0&0&0&0&0&0&0&1&0&0&0&0&0&0&0&0&0&0\\
      Windows Chrome Unidentified&0&0&.02&.09&0&0&0&0&.02&0&0&0&0&0&0&0&0&0&0&0&.86&0&0&0&0&0&0&0&0&0\\
      Ubuntu Chrome Facebook&0&0&0&0&0&0&0&.3&0&0&0&0&.04&0&0&0&0&0&0&0&0&.67&0&0&0&0&0&0&0&0\\
      Windows Firefox Unidentified&0&0&.06&0&0&0&0&0&0&0&0&0&0&0&0&0&0&0&0&0&0&0&.94&0&0&0&0&0&0&0\\
      Ubuntu Firefox Facebook&0&.06&0&0&0&0&0&0&0&.11&0&.28&0&0&0&0&0&0&0&0&0&0&0&.56&0&0&0&0&0&0\\
      OSX Chrome Twitter&0&0&0&0&0&0&0&.13&0&0&0&0&0&0&0&0&0&0&0&0&0&0&0&0&.75&0&0&0&.06&.06\\
      Windows IExplorer Unidentified&.71&0&0&0&0&0&0&0&0&0&0&0&0&0&0&0&0&0&0&0&0&0&0&0&0&.29&0&0&0&0\\
      Ubuntu Non-Browser Skype&0&0&0&0&0&0&0&1&0&0&0&0&0&0&0&0&0&0&0&0&0&0&0&0&0&0&0&0&0&0\\
      Windows IExplorer Google-Services&0&0&0&0&0&0&0&0&0&0&0&0&0&0&0&0&0&0&0&0&0&0&0&0&0&0&0&1&0&0\\
      OSX Chrome Google-Services&0&0&0&0&0&0&0&0&0&0&0&0&0&0&0&0&0&0&0&0&0&0&0&0&0&0&0&0&1&0\\
      OSX Chrome Unidentified&0&0&0&0&0&0&0&0&0&0&0&0&0&0&0&0&0&0&0&0&0&0&0&0&0&0&0&0&0&1\\
    }
  \end{minipage}
  
  %

  \caption{Confusion matrix (rows are ground truth). For most tuples
    the classification is almost perfect. Exceptions occur mostly
    between similar tuples and the Unidentified classes (which may actually
    be a correct answer that we could not verify). For example, ``Ubuntu
    Chrome Google-Services'' was mistakenly classified as ``Ubuntu
    Chrome Unidentified'' in 18\% of the cases and ``Ubuntu Firefox
    Google-Services'' in 7\%.}
  \label{fig:confusionmat}
\end{figure*}

%% file: acc_fig_new.tex
\begin{figure*}[t!]
\begin{minipage}[t]{0.4\linewidth}
\centering
\accResultsFirst{Tuple_results.tex}{102}
\subcaption{Tuple Accuracy Results}
\label{mp:tupleresults}
\end{minipage}
\hspace{20pt}
\begin{minipage}[t]{0.4\linewidth}
\centering
\accResults{results_browser_clean.tex}{102}
\vspace{-11pt}
\subcaption{Browser Accuracy Results}
\label{mp:browsers}
\end{minipage}

\vspace{10pt}
\begin{minipage}[t]{0.4\linewidth}
\centering
\accResults{results_os_clean.tex}{102}
\vspace{-11pt}
\subcaption{OS Accuracy Results}
\label{mp:oss}
\end{minipage}
\hspace{20pt}
\begin{minipage}[t]{0.4\linewidth}
\centering
\accResults{results_app_clean.tex}{102}
\vspace{-11pt}
\subcaption{Application Accuracy Results}
\label{mp:applications}
\end{minipage}

\begin{center}
\ref{barLagendName}
\end{center}
\caption{Accuracy results for KNN, SVM-RBF, SVM-MAP, SVM-SIM, RF with different features sets that are equivalent to adversarial opponents. The captions are the same for the four subfigures}
\label{fig:results}
\end{figure*}

%% file: vpn_chiper_short.tex
\begin{figure*}[htb]
\begin{minipage}[t]{0.5\linewidth}
 	\centering
	\VPNCipherFirst{vpn_os_browser.tex}{102}
  \subcaption{VPN OS and Browser}
 \label{fig:VPN_OS_Browser_short}
\end{minipage}
\hspace{40pt}
\begin{minipage}[t]{0.5\linewidth}
 	\centering
	\VPNCipher{cipher_os_browser_new.tex}{102}
  \subcaption{Cipher OS and Browser}
 \label{fig:Cipher_OS_Browser_short}
\end{minipage}
\begin{center}
\ref{barLagendNameVPN}
\end{center}
\caption{The influence of using VPN or changing the Cipher suite at test time on the accuracy results.}
\label{fig:VPN_Cipher_short}
\end{figure*}

%% file: var_training_data_set_size.tex
\begin{figure*}[htb!]
\centering
  \begin{minipage}[t]{0.3\linewidth}
	\centering
	\varsTrainSizesFirst{knn_results.tex}
	\subcaption{KNN}
	\label{fig:KNN_var}
  \end{minipage}			
  \begin{minipage}[t]{0.3\linewidth}
	\centering
	\varsTrainSizes{svmrbf_results.tex}
	\subcaption{SVM-RBF}
	\label{fig:SVM-RBF_var}
  \end{minipage}
  \begin{minipage}[t]{0.3\linewidth}
	\centering
	\varsTrainSizes{RandomForest_results.tex}
	\subcaption{RF}
	\label{fig:RF_var}
  \end{minipage}
  
  \ref{TimeLagendName2}
  \caption{Accuracy of KNN, SVM-RBF, RF with various training data set size (Number of Samples). The x-axis is in logarithmic scale.}
  \label{fig:var_data_size} 
\end{figure*}

%% file: acc_realtime.tex
\begin{figure*}[t!]
\begin{minipage}[t]{0.4\linewidth}
	\centering
	\accResultsFirst{real_time_results_10min.tex}{102}
  \subcaption{Up to 10 min on KNN, SVM-RBF, SVM-MAP, SVM-SIM, RF classification}
 \label{fig:realtime_10M}
\end{minipage}
\hspace{20pt}
\begin{minipage}[t]{0.4\linewidth}
	\centering
	\accResults{real_time_results_1min.tex}{102}
	\vspace{-11pt}
  \subcaption{Up to 1 min on KNN, SVM-RBF, SVM-MAP, SVM-SIM, RF classification}
 \label{fig:realtime_1M}
\end{minipage}

\vspace{10pt}
\begin{minipage}[t]{0.4\linewidth}
 	\centering
	\accResults{real_time_results_10sec.tex}{102}
	\vspace{-11pt}
  \subcaption{Up to 10 sec on KNN, SVM-RBF, SVM-MAP, SVM-SIM, RF classification}
 \label{fig:realtime_10S}
\end{minipage}
\hspace{20pt}
\begin{minipage}[t]{0.4\linewidth}
 	\centering
	\accResults{real_time_results_1sec.tex}{102}
	\vspace{-11pt}
  \subcaption{Up to 1 sec on KNN, SVM-RBF, SVM-MAP, SVM-SIM, RF classification}
 \label{fig:realtime_1S}
\end{minipage}
\begin{center}
\ref{barLagendName}
\end{center}
\caption{Real time Tuple classification of short period of a session (1 second, 10 seconds, 1 minute and 10 minutes) using the KNN, SVM-RBF, SVM-MAP, SVM-SIM, RF classification algorithms. The captions are the same for the four subfigures}
\label{fig:short_sessions}
\end{figure*}

%% file: acc_runtime_split.tex
\begin{figure*}[t!]
\centering
		\begin{minipage}[t]{0.45\linewidth}
		\centering
			\TrainTestFirst{knn_times_train_CHECK.tex}{450}
			\subcaption{KNN Train}
			\label{fig:KNN_runtime_train}
		\end{minipage}
		\begin{minipage}[t]{0.45\linewidth}
		\centering
			\TrainTestLog{knn_times_test_CHECK.tex}{3}
			\subcaption{KNN Test - one sample}
			\label{fig:KNN_runtime_test}
		\end{minipage}
		
		\begin{minipage}[t]{0.45\linewidth}
		\centering
			\TrainTestLog{svmrbf_times_train_CHECK.tex}{450}
			\subcaption{SVM+RBF, SVM+MAP, SVM+SIM Train}
			\label{fig:SVM_runtime_train}
		\end{minipage}
		\begin{minipage}[t]{0.45\linewidth}
		\centering
			\TrainTestLog{svmrbf_times_test_CHECK.tex}{3}
			\subcaption{SVM+RBF, SVM+MAP, SVM+SIM Test - one sample}
			\label{fig:SVM_runtime_test}
		\end{minipage}
		
		\begin{minipage}[t]{0.45\linewidth}
		\centering
			\TrainTestLog{RandomForest_times_train_CHECK.tex}{450}
			\subcaption{RF Train}
			\label{fig:RF_runtime_train}
		\end{minipage}
		\begin{minipage}[t]{0.45\linewidth}
		\centering
			\TrainTestLog{RandomForest_times_test_CHECK.tex}{3}
			\subcaption{RF Test - one sample}
			\label{fig:RF_runtime_test}
		\end{minipage}
		
		\ref{TimeLagendName2}
		\caption{Training (including cross validation) and testing
          time of the machine learning algorithms for various training
          dataset sizes. The x-axis is in logarithmic scale. The testing time for one sample (number of samples = 6190). The captions are the same for the six subfigures}
		\label{fig:runningTimeTrainTest}
\end{figure*}